# CHAOS ANTI-SYNCHRONIZATION IN MULTIPLE TIME DELAY POWER SYSTEMS

**E. M. SHAHVERDIEV**[1*]

[1]Institute of Physics, 33, H. Javid Avenue, Baku, AZ1143, Azerbaijan.

**AUTHOR'S CONTRIBUTION**
The sole author designed, analyzed and interpreted and prepared the manuscript.

*Original Research Article*

**ABSTRACT**

We elucidate conditions for chaos anti-synchronization between two uni-directionally coupled multiple time delay power systems. The results are of some importance to prevent power black-out in the entire power grid.

## 1 Introduction

Chaos synchronization [1-7] is one of the basic features in nonlinear science and it is one of fundamental importance in a variety of complex physical, chemical, power and biological systems with possible application areas in secure communications, optimization of nonlinear system performance, modeling brain activity and pattern recognition phenomena, avoiding power black-out.

Finite signal transmission times, switching speeds and memory effects makes time-delayed systems ubiquitous in nature, technology and society [8]. Therefore the study of synchronization phenomena in such systems is of considerable practical importance. Some power systems are also described by time delay systems, see e.g. [9-10] and references there-in.

Electrical power systems are essentially nonlinear dynamical systems [11-14]. Study of chaos and its control in such systems could be of considerable importance from the point of view of avoiding undesirable behaviors such as power black-out. Synchronization in power systems is of huge importance from the management point of view of complex power systems.

This paper is devoted to chaos anti-synchronization in some simple power models- a single -machine-infinite-bus systems (SMIB) power system [10,11-14]. We study chaos anti-synchronization in power systems described by both linearly and non-linearly coupled multiple time delay power systems and elucidate conditions for such synchronization. To the best of our knowledge this is the first report on anti-synchronization between two power systems with multiple time delays. Additionally, research in chaos synchronization regimes in multiple time delay systems is of some importance for improved security in chaos based communication systems.

___

*Corresponding author: shahverdiev@physics.ab.az;

This paper is organized as follows. In section II we briefly discuss power black-out and present the essence of the approach by which power black-out can be avoided in the context of synchronization and anti-synchronization. Section III deals with the anti-synchronization behavior in multiple time delay power systems. Section IV presents the results of numerical simulations of the model considered in Section III. Conclusions are given in Section V.

## 2 Power Black-Out, Synchronization and Anti-Synchronization

A power black-out is a short- or long-term loss of the electric power to an area [15]. There are many causes of power failures in an electricity network. Usually it is assumed that synchronization between different parts facilitates flawless functioning of the entire system. Let us assume that some parts of the power system fail to function properly and local power black-out occurs [3, 16-18]. A natural response to avoid the total power black-out throughout the entire system could be to achieve anti synchronization between the failed part and the rest of the power system.

We underline that earlier an anti-synchronization regime was proposed in [19] to avoid the danger of global extension of species. Recently such an approach based on synchronization and anti-synchronization transition is also proposed to avoid the spread of some infectious deceases, see e.g. [20] and references there-in.

The results of the investigation carried out in this paper underscore the principal possibility of anti-synchronization between the power systems coupled both linearly and nonlinearly. This testifies to the wide range applicability of the results in the practical situations. The main conclusion of the paper is the possibility of avoiding possible power black-out in the power grid by anti-synchronizing different parts of the grid.

## 3 Chaos Anti-Synchronization in Multiple Time Delay Power Systems

Consider the classical SMIB power system [10, 11-14], whose phase dynamics is described by the swing equations generalized for two time delays

$$M \frac{d\theta^2}{dt^2} + D \frac{d\theta}{dt} + P_{\max} \sin\theta = P_m + P_{\tau_1} + P_{\tau_2}, \quad (1)$$

which can be written as a system of first order equations:

$$\frac{dx_1}{dt} = x_2 \quad (2)$$

$$\frac{dx_2}{dt} = -cx_2 - \beta \sin x_1 + f \sin \varpi t + k_1 \sin(R_1 x_1(t-\tau_1)) + k_2 \sin(R_2 x_1(t-\tau_1)) \quad (3)$$

Where

$$x_1 = \theta, x_2 = \frac{d\theta}{dt}, c = \frac{D}{M}, \beta = \frac{P_{\max}}{M}, f = \frac{A}{M} \quad (4)$$

Here $\theta$ is the phase of the oscillator; $M$ is the moment of inertia; $D$ is the damping constant; $P_m$ is the power of the machine; $P_{\max}$ is the maximum power of generator;

$$P_m = A\sin \omega t; P_{\tau_1} = m_1 \sin(R_1 x_1(t-\tau_1)) \text{ and } P_{\tau_2} = m_2 \sin(R_2 x_1(t-\tau_2))$$

are the delayed feedback terms with $k_1 = m_1/M$ and $k_2 = m_2/M$; $R_1$ and $R_2$ are frequency modulation of the feedback terms $x_1(t-\tau_1))$ and $x_2(t-\tau_1))$ respectively; $k_1$ and $k_2$ are the feedback strengths.

Introducing a new state variable via $x_3 = \sin \omega t$ this SMIB power system can be written as a system of three differential equations without variable time term in the right-hand side:

$$\frac{dx_1}{dt} = x_2 \tag{5}$$

$$\frac{dx_2}{dt} = -cx_2 - \beta \sin x_1 + f \sin x_3 + k_1 \sin(R_1 x_1(t-\tau_1)) + k_2 \sin(R_2 x_1(t-\tau_1)) \tag{6}$$

$$\frac{dx_3}{dt} = \omega \tag{7}$$

Extensive investigation of dynamical behavior of this power system in connection with complete synchronization has been conducted in [10]. In this paper our task is to investigate chaos anti-synchronization between linearly and non-linearly coupled multiple time delay power systems, e.g. $x_2 = -y_2$. For that purpose first we consider the following delay differential equations (DDE) system coupled with the system of Eqs. (5-7) by the coupling term $K(-x_2 - y_2)$:

$$\frac{dy_1}{dt} = y_2 \tag{8}$$

$$\frac{dy_2}{dt} = -cy_2 - \beta \sin y_1 + f \sin y_3 + k_3 \sin(R_1 y_1(t-\tau_1)) + k_4 \sin(R_2 y_1(t-\tau_1)) + K(-x_2 - y_2) \tag{9}$$

$$\frac{dy_3}{dt} = \omega \tag{10}$$

In deriving the existence conditions for the anti-synchronization regime we heavily benefited from the approach developed in our previous papers, which were devoted to synchronization regimes in popular laser physics Ikeda and Lang-Kobayashi models [21-22]. In the case of linear coupling between the systems the existence condition for the anti-synchronization regime $x_2 = -y_2$ is of the following form:

$$k_1 = k_3, k_2 = k_4 \tag{11}$$

Indeed in order to obtain this result we introduce the error signals:

$$\Delta_1 = x_1 + y_1, \Delta_2 = x_2 + y_2 \tag{12}$$

Next from Eqs.(7) and (10) with some generalization we easily establish that $x_3 = \omega t + \phi_x$, $y_3 = \varpi t + \phi_y$, where $\phi_x$ and $\phi_y$ are the initial phases for $x_3$ and $y_3$, respectively.

Then we add Eqs.(5) and (8), and Eqs.(6) and (9). For *small* error signals under conditions $k_1 = k_3, k_2 = k_4$ we obtain that

$$\frac{d\Delta_1}{dt} = \Delta_2 \tag{13}$$

$$\frac{d\Delta_2}{dt} = -(c+K)\Delta_2 - \beta \Delta_{1,\tau_1} \cos x_1 + k_1 R_1 \Delta_{2,\tau_1} \cos(R_1 x_{1,\tau_1}) + k_2 R_2 \Delta_{2,\tau_2} \cos(R_2 x_{1,\tau_2}) \tag{14}$$

For brevity in Eq.(14) it is accepted that e.g. $\Delta_{1,\tau_1} \equiv \Delta_1(t - \tau_1)$, $x_{1,\tau_2} \equiv x_1(t - \tau_2)$.

We also note that without loss of generality while adding Eqs.(6) and (9) the sum $f(\sin x_3 + \sin y_3) = f(\sin(\varpi t + \phi_x) + \sin(\varpi t + \phi_y))$ can be nullified under time independent conditions $\phi_x - \phi_y = \pi + 2\pi n$ with $n$ being an integer number. For completeness we also present time dependent condition: $2\varpi t + \phi_x + \phi_y = 2\pi n$.

It is clear that $\Delta_1 = 0$ and $\Delta_2 = 0$ satisfy Eqs.(13) and (14) under assumed conditions (11). Conditions (11) are the existence conditions for the anti-synchronization regime in the case of linear coupling. Unfortunately for high dimensional time delay systems finding the stability conditions for the synchronization regimes studied in this paper is rather difficult.

Below we will also consider the case of nonlinearly coupled power systems. For this case Eq.(9) can be written as

$$\frac{dy_2}{dt} = -cy_2 - \beta \sin y_1 + f \sin y_3 + k_3 \sin(R_1 y_1(t-\tau_1)) + k_4 \sin(R_2 y_1(t-\tau_1)) + k_5 \sin(R_1 x_1(t-\tau_3)), \tag{15}$$

In other words the linear coupling term $K(-x_2 - y_2)$ is replaced by the nonlinear coupling term $k_5 \sin(R_1 x_1(t - \tau_3))$.

Here $\tau_3$ is the coupling delay time between the power systems; $k_5$ is the coupling strength between the systems. Existence conditions for anti-synchronization can be established as it was found for the linearly coupled systems. Indeed under conditions

$$k_3 = k_1 + k_5, k_2 = k_4 \qquad (16)$$

and restrictions imposed on the initial phases derived above Eq.(14) ((Eq.13) remains unchanged) can be written as:

$$\frac{d\Delta_2}{dt} = -c\Delta_2 - \beta\Delta_{1,\tau_1} \cos x_1 + k_1 R_1 \Delta_{2,\tau_1} \cos(R_1 x_{1,\tau_1}) + k_2 R_2 \Delta_{2,\tau_2} \cos(R_2 x_{1,\tau_2}) \qquad (17)$$

While obtaining Eq.(17) we put $\tau_3 \equiv \tau_1$. Clearly $\Delta_1 = 0$ and $\Delta_2 = 0$ satisfy Eqs.(13) and (17). It is worth mentioning that in case of nonlinear coupling there is an additional anti-synchronization regime $x_2 = -y_2$ under different existence conditions:

$$k_1 = k_3, k_4 = k_2 + k_5. \qquad (18)$$

Indeed under conditions (18), with $\tau_3 \equiv \tau_2$ and the above mentioned initial phase restrictions, Eq.(17) takes the following form:

$$\frac{d\Delta_2}{dt} = -c\Delta_2 - \beta\Delta_{1,\tau_1} \cos x_1 + k_1 R_1 \Delta_{2,\tau_1} \cos(R_1 x_{1,\tau_1}) + k_4 R_2 \Delta_{2,\tau_2} \cos(R_2 x_{1,\tau_2}) \qquad (19)$$

Again $\Delta_1 = 0$ and $\Delta_2 = 0$ satisfy Eqs. (13) and (19).

As derivation of the stability conditions for the anti-synchronization regime is difficult, below we also present the results of numerical simulations for both linearly and non-linearly coupled systems. We emphasize that in numerical simulations we choose the coupling delay time between the systems $\tau_3$ different from $\tau_1$ and $\tau_2$. As presented here (see Figs.2 and 3) despite this we achieve high value for the correlation coefficient, which testify to the stability of the anti-synchronization regime under the parameter mismatches. Detailed study of the effect of multi-parameter influence on the correlation coefficient is beyond the scope of this paper and will be reported elsewhere. We also note that in the case of nonlinearly coupled double time delay systems the number of possible synchronization regimes is increased twofold.

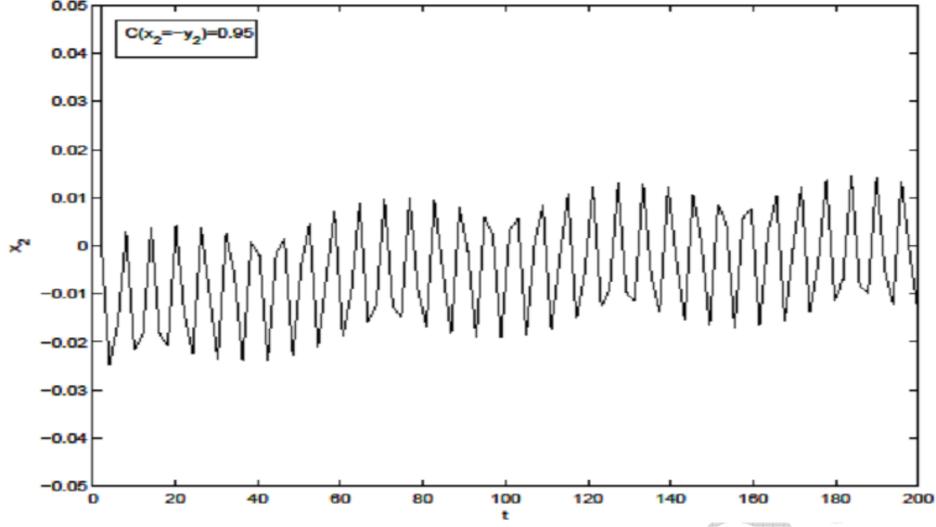

**Fig. 1.** Numerical simulation of the linearly coupled multiple time delay power models, Eqs.(5-7) and Eqs.(8-10):time series of $x_2$ is shown. $C$ is the correlation coefficient between the power systems: $x_2 = -y_2$. the parameters are

$c = 65, \beta = 1, f = 1, \varpi = 1, R_1 = R_2 = 0.005, k_1 = k_3 = 0.2, k_2 = k_4 = 0.4,$
$\tau_1 = 0.6, \tau_2 = 0.2, K = 100.$

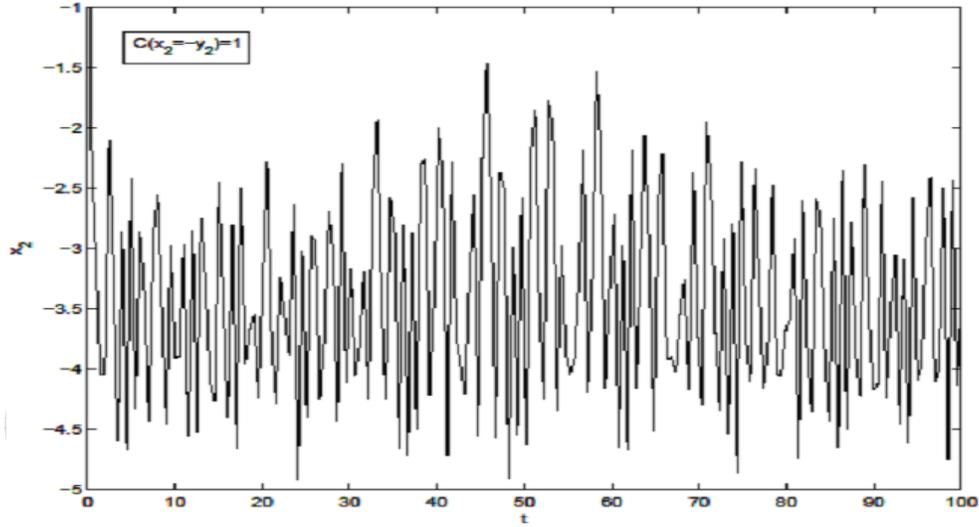

**Fig. 2.** Numerical simulation of the non-linearly coupled multiple time delay power models, Eqs.(5-7) and Eqs.(8,15,10):time series of $x_2$ is shown. $C$ is the correlation coefficient between the power systems $x_2 = -y_2$. the parameters are

$c = 1, \beta = 1, f = 1, \varpi = 1, R_1 = R_2 = 5, k_3 = 20, k_1 = 0.1,$
$k_5 = 19.9, k_2 = k_4 = 0.4, \tau_1 = 0.1, \tau_2 = 0.2, \tau_3 = 0.3$

## 4 Results and Discussions

Now we demonstrate that two multiple time delay coupled power systems can be anti-synchronized. We use DDE software built-in Matlab 7 for the numerical simulations. We study anti-synchronization between the power systems using the correlation coefficient $C$ [23]

$$C(\Delta t) = \frac{<(x(t)-<x>)(y(t+\Delta t-<y>)>}{\sqrt{<(x(t)-<x>)^2><(y(t+\Delta t)-<y>)^2>}} \tag{20}$$

where $x$ and $y$ are the outputs of the interacting systems; the brackets $<.>$ represent the time average; $\Delta t$ is a time shift between the systems outputs: for our case $\Delta t = 0$. This coefficient indicates the quality of synchronization: $C = -1$ means perfect anti-synchronization: $C(x=y) = -1$ or equivalently $C(x=-y) = 1$.

Fig.1 demonstrates the dynamics of variable $x_2$ for Eqs.(5-7) and $C$ shows the quality of anti-synchronization between the linearly coupled multiple time delay power systems, Eqs.(5-7) and Eqs.(8-10) ($x_2 = -y_2$). The parameters are

$$c = 65, \beta = 1, f = 1, \varpi = 1, R_1 = R_2 = 0.005, k_1 = k_3 = 0.2, k_2 = k_4 = 0.4,$$
$$\tau_1 = 0.6, \tau_2 = 0.2, K = 100.$$

Fig.2 depicts the dynamics of variable $x_2$ for system of Eqs. (5-7) with $C$ indicating the level of anti-synchronization between the nonlinearly coupled power systems, Eqs.(5-7) and Eqs.(8,15,10). The parameters are

$$c = 1, \beta = 1, f = 1, \varpi = 1, R_1 = R_2 = 5, k_3 = 20, k_1 = 0.1,$$
$$k_5 = 19.9, k_2 = k_4 = 0.4, \tau_1 = 0.1, \tau_2 = 0.2, \tau_3 = 0.3$$

with existence condition of the anti-synchronization regime $k_3 = k_1 + k_5$ and $k_2 = k_4$.

In Fig.3 we present another example of anti-synchronization between the nonlinearly coupled systems with existence conditions $k_1 = k_3$ and $k_4 = k_2 + k_5$:

$k_1 = k_3 = 0.2, k_4 = 15, k_2 = 0.1, k_5 = 14.9$; the other parameters are:

$$c = 1, \beta = 1, f = 1, \varpi = 1, R_1 = R_2 = 5, \tau_1 = 0.1, \tau_2 = 0.2, \tau_3 = 0.3$$

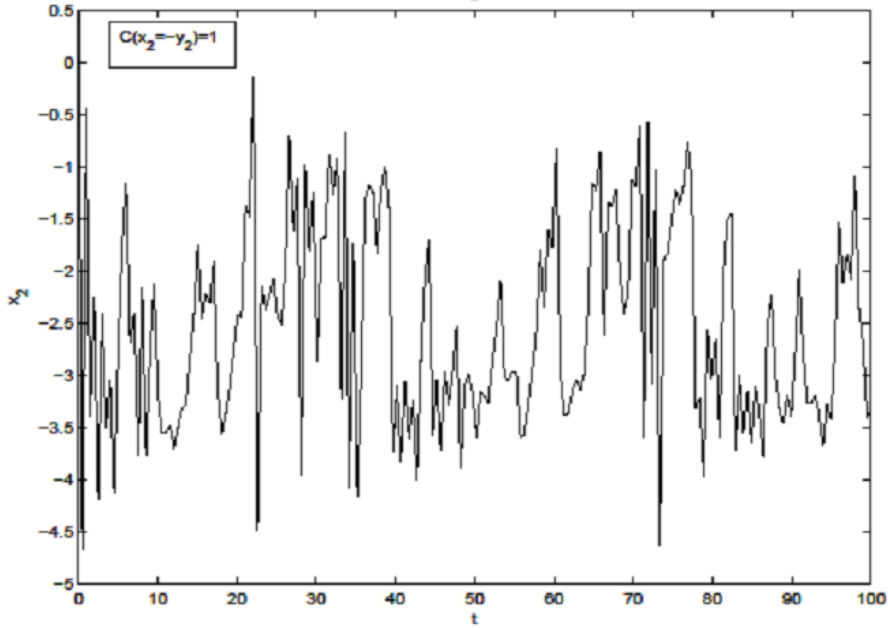

**Fig. 3. Numerical simulation of the non-linearly coupled multiple time delay power models,Eqs.(5-7) and Eqs.(8,15,10):time series of $x_2$ is shown. $C$ is the correlation coefficient between the power systems: $x_2 = -y_2$. The parameters are**

$$c=1, \beta=1, f=1, \varpi=1, R_1=R_2=5, \tau_1=0.1, \tau_2=0.2, \tau_3=0.3$$
$$k_1=k_3=0.2, k_4=15, k_2=0.1, k_5=14.9.$$

## 5 Conclusions

The results of the investigation show the possibility of anti-synchronization between the power systems coupled both linearly and nonlinearly. This may be of some importance in the practical situations. The main application could be in the field of avoiding possible power black-out in the power grid by anti-synchronizing different parts of the grid. The approach can be used for the instability control in the power grid.

We have studied the simplest power system in the parameter domain which generated chaotic behavior. Chaos is the only one of the possible solutions. There are also possible non-chaotic behavior such as periodic or quasi-periodic solutions for some parameter values. Although we have presented the results for the chaotic anti-synchronization behavior, the conclusion is equally applicable to the case of non-chaotic systems.

## Acknowledgements

This research was supported by a Marie Curie Action (return phase) European Community Framework Programme. The author gratefully acknowledge Prof. K.A. Shore (Bangor University, UK) for comments on the manuscript.

## Competing Interests

Author has declared that no competing interests exist.